\documentclass[twocolumn,prl]{revtex4}
\usepackage{graphicx}
\usepackage{bm}
\usepackage{amsmath}
\begin{document} \title{Screening by composite charged particles: the case of quantum well trions}

\author{C. Ciuti and G. Bastard}
\affiliation{Laboratoire Pierre Aigrain, Ecole Normale
Sup\'erieure, 24, rue Lhomond, 75005 Paris, France }

\begin{abstract} We study the screening of an external potential
produced by a two-dimensional gas of charged excitons (trions). We
determine the contribution to the dielectric function induced by
these composite charged particles within a random phase
approximation. In mixtures of free electrons and trions, the trion
response is found dominant. In the long wave-length limit, trions
behave as point charges with mass equal to the sum of the three
particle components. For finite wave-vectors, we show how the
dielectric response is sensitive to the composite nature of trions
and the internal degrees of freedom.  Predictions are presented
for the screening of a Coulomb potential, the scattering by
charged impurities and the properties of trionic plasmons.
\end{abstract}

\date{\today} \maketitle

A major breakthrough in our understanding of the optical response
of doped semiconductor quantum wells (QWs) was achieved when the
existence of trions was demonstrated experimentally both in II-VI
and III-V semiconductors
\cite{discovery,Finkelstein_1995,Shields_1995}. These composite
particles appear as the weakly bound state of an exciton (a bound
electron-hole pair) to an electron or a hole depending on the
n-type or p-type doping of the QW. For the sake of clarity, in the
following we will deal with negatively charged excitons ($X^-$).
The physics of these composite charged particles in the low
excitation regime has attracted a considerable interest,
especially concerning the metal-insulator transition
\cite{Finkelstein_1995,Eytan_1998}, the intrinsic radiative
recombination efficiency \cite{Esser_2000,Ciulin_2000}, the
singlet-triplet crossover under strong magnetic
field\cite{Yusa_2001}, the role of phonons in the diffusion
properties \cite{Portella_2002}, and the remarkable drift
transport induced by an applied electric
bias\cite{Sanvitto_2001,Pulizzi_2003}.

More recently, the first investigations of a dense trion gas have
been performed by resonant optical pump-probe\cite{Portella_2003}.
In principle, by optical pumping of the trion resonance, it should
be possible to "convert" partially or even completely the
background electron gas into a trion gas. Hence, it can be
possible to study the many-body properties of a gas of composite
charged particles. One interesting issue to be addressed is the
screening response of such a peculiar gas to an external
potential, such as the Coulomb field generated by a charged
impurity. In this respect, the description of a trion gas as point
particles (like electrons) should be only a long wave-length
approach, while at shorter wave-length the trion granularity
should show up. Since the impurity-induced scattering, the dc
mobility and the collective excitations are sensitive to different
wave-lengths, it is desirable to determine the complete response
of a dense trion gas to external disturbances and to check whether
and when the internal structure can be important. In this letter,
we will present a Random Phase Approximation (RPA) treatment of
the trion gas response. We will show the role of the composite
nature of trions in the determination of the dielectric response.

To simplify our description, we shall present results for a purely
two-dimensional system, omitting the form factors due to the
finite extension of the quantum well wave-functions of electrons
and holes along the QW growth direction. An $X^-$ trion state is
represented by a three-body wave-function, in which the center of
mass, the relative motion and the spin part can be factorized,
namely
\begin{equation}
\Psi_{{\bm k},\eta,S_e,S_{hz}} = \frac{e^{i {\bm k} \cdot {\bm
R}}}{\sqrt{A}}~  ~ \phi_{\eta,S_e}({\bm \lambda}_1,{\bm
\lambda}_2)~\chi_{S_e,S_{hz}}(s_{1z},s_{2z},s_{hz}),
\end{equation}
where ${\bm k}$ and ${\bm R}$ are the center of mass wave-vector
and position respectively, $A$ is the sample area, $\eta$ is the
relative motion quantum number, while $S_e$ and $S_{hz}$ denote
the spin state. The relative motion part depends on the variables
${\bm {\lambda}}_1 = {\bm \rho}_1 - {\bm \rho}_h$ and ${\bm
\lambda}_2= {\bm \rho}_2 - {\bm \rho}_h$, where ${\bm \rho}_1$,
${\bm \rho}_2$, ${\bm \rho}_h$ are the in-plane positions of the
first, second electron and the hole respectively, while $s_{1z}$,
$s_{2z}$, $s_{hz}$ are the spin components along the perpendicular
direction $z$. In semiconductor quantum wells, heavy-hole trions
have eight spin states, where $S_e \in \{0,1 \}$ represents the
total spin of the two electrons and $S_{hz} \in \{\pm 3/2\}$ is
the heavy-hole band angular momentum projection. The two states
with $S_e = 0$ are called singlet trions, while the six states
with $S_e= 1$ are triplet trions. To fulfil the Pauli exclusion
principle, for a singlet (triplet) spin state, the relative motion
wave-function $\phi_{\eta,S_e}$ is symmetric (anti-symmetric)
under exchange of the positions of the two electrons. The $X^-$
charge density operator reads
\begin{equation}
\hat{n}({\bm \rho}) = - e \left (\delta^{(2)}({\bm \rho}-{\bm
\rho_1})+\delta^{(2)}({\bm \rho}-{\bm \rho_2})-\delta^{(2)}({\bm
\rho}-{\bm \rho_h})\right )~.
\end{equation}
The matrix elements of $\hat{n}({\bm \rho})$ on the basis of the
trion states are diagonal with respect to the spin indexes. They
can be written as
\begin{equation}
\langle {\bf k'}, \eta',S_e,S_{hz}|\hat{n}| {\bf k},
\eta,S_e,S_{hz} \rangle = - \frac{e}{A}~e^{i ({\bm k}-{\bm k'})
\cdot {\bm \rho}}~ \mathcal{T}^{\eta',\eta,S_e}_{{\bm k}-{\bm
k'}}~,
\end{equation} where the trion "granularity" factor $\mathcal{T}^{\eta',\eta,S_e}_{\bm q}$
accounting for the composite nature is
\begin{equation}
\mathcal{T}^{\eta',\eta,S_e}_{\bm q} = \alpha_{e,{\bm
q}}^{\eta',\eta, S_e} + \alpha_{e,-{\bm q}}^{\eta',\eta,S_e} -
\alpha_{h,{\bm q}}^{\eta',\eta,S_e},
\end{equation}
where the electron contribution reads
\begin{equation}
\alpha_{e,{\bm q}}^{\eta',\eta,S_e} = \int  d^2 {\bm \lambda_1}
d^2 {\bm \lambda_2} ~\phi_{\eta',S_e}^{~\star} ~ \phi_{\eta,S_e} ~
e^{i {\bm q} \cdot \left ( \frac{m_e}{M} {\bm \lambda_2}-
\frac{m_e+m_h}{M} {\bm \lambda_1} \right  )}~,
\end{equation}
and the hole part is
\begin{equation}
\alpha_{h,{\bm q}}^{\eta',\eta,S_e} = \int  d^2 {\bm \lambda_1}
d^2 {\bm \lambda_2} ~\phi_{\eta',S_e}^{~\star} ~ \phi_{\eta,S_e} ~
e^{i {\bm q} \cdot \left (\frac{m_e}{M} ({\bm \lambda_1}+{\bm
\lambda_2}) \right  )}~,
\end{equation}
with $m_e$, $m_h$, $M = 2m_e + m_h$ the electron, hole and trion
masses respectively. If an external potential $V_{\text{ext}}({\bm
\rho},t) = \left (\sum_{{\bm q}} \frac{1}{2} V_{\text{ext}}({\bm
q},\omega) e^{i({\bm q}\cdot {\bm \rho}-\omega t)} + h.c. \right
)$ acts on the system, then the trionic wave-functions are
perturbed. The perturbation of the trion wave-function generates a
non-homogeneous charge density, which creates a local Hartree
potential and an exchange-correlation correction (we neglect it
here). Hence, the screened potential is given by the
self-consistent equation $V_{\text{s}} = V_{\text{ext}} +
V_{\text{loc}}$, where $V_{\text{loc}}({\bm \rho},t) = \int d^2
{\bm \rho'} \frac{e}{\kappa |{\bm \rho}-{\bm \rho '}|}~\delta
n({\bm \rho'},t)$, being $\delta n({\bm \rho},t)$ the induced
charge density and $\kappa$ is the static dielectric constant of
the semiconductor quantum well.
 The self-consistent potential
energy felt by a trion is
\begin{equation}
U_{\text{s}}^{\text{tr}}({\bm \rho_1},{\bm \rho_2},{\bm \rho_3},t)
= -e \left ( V_{\text{s}}({\bm \rho_1},t) + V_{\text{s}}({\bm
\rho_2},t) - V_{\text{s}}({\bm \rho_h},t) \right )~.
\end{equation}
The matrix elements of the perturbation energy on the trion states
at $t=0$ are
\begin{equation}
\langle {\bf k'}, \eta',S_e,S_{hz} |U_{\text{s}}^{\text{tr}} |
{\bf k}, \eta,S_e,S_{hz} \rangle =  -e~ V_{\text{s}}({\bm k}-{\bm
k'},\omega)~\mathcal{T}^{\eta',\eta,S_e}_{{\bm k}-{\bm k'}}~.
\end{equation}
By calculating the lowest-order perturbation theory for the trion
wave-functions and summing incoherently the contribution from all
the populated trion states (RPA), we get the the screened
potential $ V_{\text{s}}({\bm q},\omega) = V_{\text{ext}}({\bm
q},\omega)/\epsilon_{X^-}({\bm q},\omega)$, where the
trion-induced dielectric function reads
\begin{equation}
\epsilon_{X^-}({\bm q},\omega)=  1-\frac{2\pi e^2}{\kappa q A}
\Pi_{X^-}({\bm q},\omega)~. \label{epsilon_trion}
\end{equation}
The trion RPA-polarization contribution is
\begin{equation}
\Pi_{X^-}({\bm q},\omega) = \sum \frac{(f_{{\bm k}+{\bm
q},\eta',S_e,S_{hz}} - f_{{\bm k},\eta,S_e,S_{hz}})
|\mathcal{T}^{\eta,\eta',S_e}_{\bm q}|^2}
 {E_{{\bm k} + {\bm q},\eta',S_e} - E_{{\bm
k},\eta,S_e} - \hbar \omega -i0^+}~, \label{bubble}
\end{equation}
where the sum is meant over ${\bm k}$, $\eta$, $\eta'$, $S_e$,
$S_{hz}$. The quantity $f_{{\bm k},\eta,S_e,S_{hz}}$ is the trion
occupation number (not necessarily at equilibrium) and $E_{{\bm
k},\eta,S_{e}}$ the orbital energy of the unperturbed trion state.
Note that for a (spin-unpolarized) electron gas
\begin{equation}
\Pi_{e^-}({\bm q},\omega) = 2 \sum_{{\bm k}} \frac{f^{(e)}_{{\bm
k}+{\bm q}} - f^{(e)}_{{\bm k}}}
 {\frac{\hbar^2 |{\bm k} + {\bm q}|^2}{2m_e}- \frac{\hbar^2 k^2}{2m_e} - \hbar \omega
 -i0^+}
\end{equation}
and $\epsilon_{e^-}({\bm q},\omega)=  1-\frac{2\pi e^2}{\kappa q
A} \Pi_{e^-}({\bm q},\omega)$. In presence of a mixed gas of
trions and electrons, the linear susceptibilities of the two
components add up, i.e. the polarization of the mixture is
$\Pi_{mix}({\bm q},\omega) = \Pi_{X^-}({\bm q},\omega) +
\Pi_{e^-}({\bm q},\omega)$.
\begin{figure}
\includegraphics[width=8.5cm]{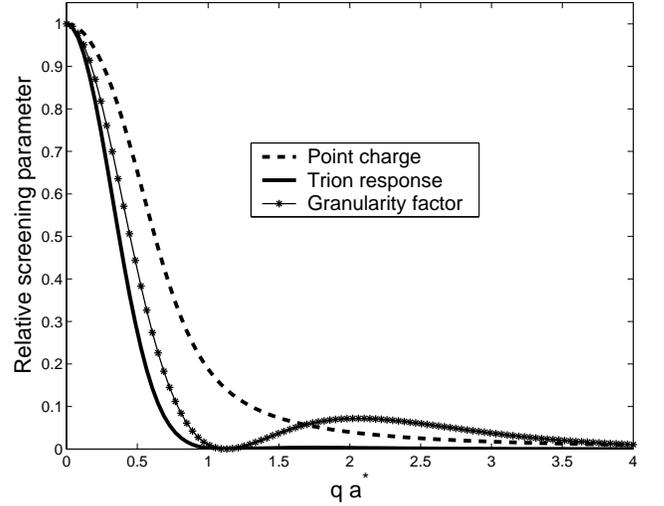}
\caption{Solid line: normalized static screening wave-vector
$q_\text{s}(q,0)/q_\text{s}(0,0)$ versus dimensionless wave-vector
$qa^{*}$ for a pure gas of trions. Dashed line: same quantity, but
without accounting for the trion granularity. Stars: the
granularity factor $|\mathcal{T}(q)|^2 =
|2\alpha_e(q)-\alpha_h(q)|^2$. Parameters: $T = 10~\text{K}$,
$a^{*} = 8~\text{nm}$, $m_e = 0.1~m_0$, $m_h = 0.2~m_0$,
$n_\text{trions} = 1 \times 10^{10}\text{cm}^{-2}$, $\kappa = 9$.
\label{fig_granularity}}
\end{figure}
Considering only the contribution of the ground trion state, the
general expression in Eq. (\ref{bubble}) can be considerably
simplified. This approximation holds when only the ground state is
populated and in the limit of small $\hbar \omega$. In the case of
zero magnetic field, the ground state is a singlet 1s-like state,
which is twice degenerate (due to the hole spin degree of
freedom). By considering only the contribution of the 1s-state, we
get
\begin{equation}
\Pi_{X^-}({\bm q},\omega) \simeq 2 \sum_{{\bm k}} \frac{(f_{{\bm
k}+{\bm q},1s} - f_{{\bm k},1s}) |\mathcal{T}(q)|^2}
 {\frac{\hbar^2 |{\bm k} + {\bm q}|^2}{2M}- \frac{\hbar^2 k^2}{2M} - \hbar \omega -i0^+}~.
\label{bubble_trion}
\end{equation}
In the following, we will consider the following trial function
\begin{equation}
\phi_{1s}({\bm \lambda}_1,{\bm \lambda}_2) = \frac{1}{2 \pi
(a^{\star})^2} \exp \left (-\frac{\lambda_1 + \lambda_2}{2
a^{\star}} \right ),
\end{equation}
where $a^{\star}$ is the effective trion radius. Accurate
numerical solutions for the internal motion of quantum well trions
are reported in the literature (see e.g. Refs.
\onlinecite{Esser_2000,binding_brasil}). With our model
wave-function, the granularity factor is $\mathcal{T}(q) =
2\alpha_e(q)-\alpha_h(q)$ with
\begin{equation}
\alpha_e(q) = \frac{1} {\left \{ \left [1+\left (\frac{m_e+m_h}{M}
~qa^{\star} \right)^2 \right] \left [ 1+ \left (\frac{m_e}{M}
~qa^{\star}\right )^2 \right ] \right \}^{3/2}}
\end{equation}
and
\begin{equation}
\alpha_h(q) = \frac{1} {\left \{ \left [ 1+ \left (\frac{m_e}{M}
~qa^{\star}\right )^2 \right ] \right \}^{3}}~.
\end{equation}

In the long wave-length limit ($q \to 0$), $|\mathcal{T}(q)|^2 \to
1$, i.e. the granularity factor does not play any role. Hence, for
small wave-vectors $q$, $X^-$ trions behave as point particles
with charge $-e$ and mass $M = 2m_e + m_h$, as expected. As shown
in Fig. \ref{fig_granularity}, the granularity factor decreases
monotonically with increasing $q$ and then has a node for the
wave-vector $\bar{q}$ such that $2 \alpha_e(\bar{q}) =
\alpha_h(\bar{q})$. This node of the granularity factor is due to
the compensation between the contributions of two electrons and
the hole within the bound $X^-$. This kind of cancellation effect
is absent in a plasma of uncorrelated electrons and holes, where
the screening contributions of the two components add up. Note
that the value of $\bar{q} a^{\star}$ depends on the mass ratio
$m_e/m_h$ and on the shape of the internal motion wave-function
(for our model wave-function $\bar{q} a^{\star} \approx 1.1$ when
$m_e/m_h =0.5$). Finally, in the short wave-length limit ($q
>> 1/a^{\star}$), the granularity asymptotically vanishes, because
the perturbing potential oscillates too quickly in the length
scale of the trion internal motion wave-function.

As a first illustrative example of our theory, we consider the
trion response to a Coulomb potential induced by a charged
impurity, that is
 $V_{\text{ext}}(q) = 2\pi e
\exp{(-qd)}/(\kappa A q)$, where $d$ is the distance between the
remote impurity donor and the quantum well plane ($d \neq 0$ for a
modulation-doped sample). The screened potential can be written as
\begin{equation}
V_{\text{s}}(q,0) = \frac{2\pi e \exp{(-qd)}}{\kappa A (q+
q_{\text{s}}(q,0))}~,
\end{equation}
being $q_{\text{s}}(q,0) = (\epsilon_{\text{mix}}(q,0)-1) q$ and
$\epsilon_{\text{mix}}$ the dielectric function for the mixed gas
of trions and electrons. In Fig. \ref{fig_granularity}, we show
the relative screening parameter
$q_{\text{s}}(q,0)/q_{\text{s}}(0,0)$ versus q (in units of the
trion effective radius $a^{\star}$) for a pure trion gas, which
can be obtained by resonant optical pumping. For simplicity, we
have taken an equilibrium Fermi-Dirac distribution function with
temperature $T = 10 K$ and with density $n_{\text{tr}} = 10^{10}
\text{cm}^{-2}$. We have used typical parameters for a CdTe-based
semiconductor\cite{discovery}, namely $m_e = 0.1 ~m_0$, $m_h =
0.2~ m_0$, $\kappa = 9$ (working in the CGS system) and a trion
radius $a^{\star} = 8 ~\text{nm}$. The thick solid line represents
$q_{\text{s}}(q,0)/q_{\text{s}}(0,0)$ due to the trion gas, while
the dashed line shows the same quantity without including the
granularity factor $|\mathcal{T}(q)|^2$. Indeed, the composite
nature of trions has a major impact in their screening response,
dramatically quenching the response at finite wave-vectors.

\begin{figure}[t!]
\includegraphics[width=8.7cm]{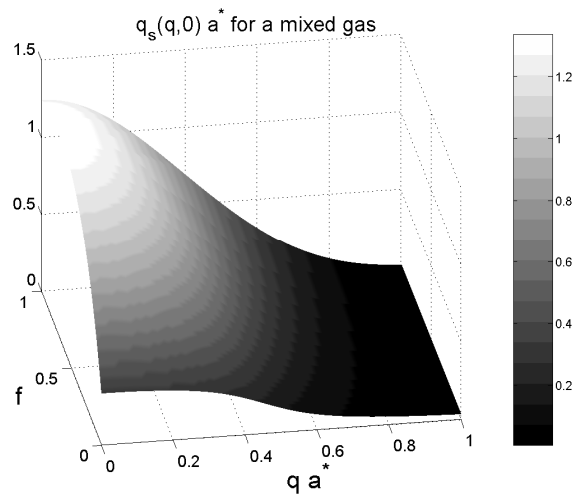}
\caption{Contours of the screening wave-vector $q_\text{s}(q,0)$
(units of $1/a^{\star}$) as a function of $qa^{*}$ and the trion
density fraction $f$ ($f = 1$ corresponds to a pure trion gas,
$f=0$ is for a pure electron gas). Parameters: total density
(electrons + trions) $n_{tot} = 10^{10}\text{cm}^{-2}$. Other
parameters as in Fig. \ref{fig_granularity}.
 \label{fig_q_fraction}}
\end{figure}

One important issue to verify is the behavior of the static
screening in presence of a mixed population of trions and
electrons. In Fig. \ref{fig_q_fraction}, we show a surface plot of
$q_{\text{s}}(q,0) a^{\star}$ as a function of the trion density
fraction $f$, which is defined by the relation $n_{\text{tr}} = f
(n_e + n_{\text{tr}}) = f n_{\text{tot}}$. Passing from a pure
electron gas ($f = 0$) to a pure trion gas ($f=1$), the static
screening changes considerably. For $q \to 0$, the screening
wave-vector increases for increasing trion fraction due to the
heavier mass of the trion. In fact, in the low temperature limit
$q_s(0,0) \propto m_e$ for a pure 2D electron gas\cite{Ando_RMP}
and therefore $q_s(0,0) \propto (2m_e+m_h)$ for a pure 2D trion
gas. Fig. \ref{fig_q_fraction} shows that even for small trion
fractions, the screening of the mixture is dominated by the trion
component. Hence, the trion granularity effects at finite
wave-vectors shown in Fig. \ref{fig_granularity} are important
also in the mixture case.

\begin{figure}[t!]
\includegraphics[width=8.7cm]{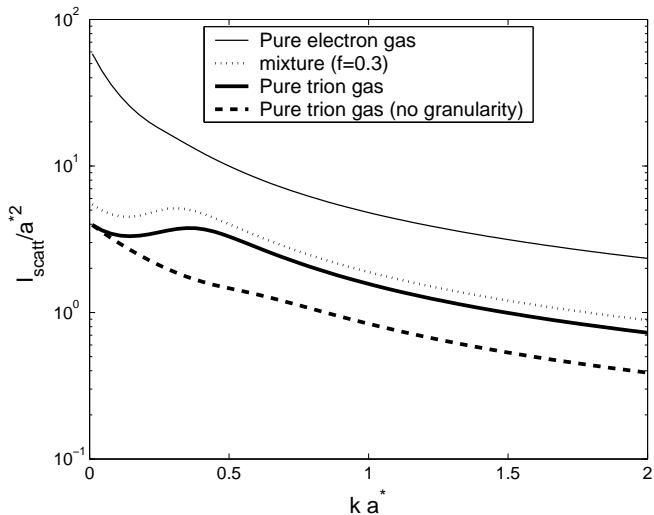}
\caption{Impurity scattering rate integral $I_{\text{int}}$ (units
of $a^{\star~2})$ as a function of $qa^{*}$. Thick solid line:
pure trion gas ($f =1$). Thick dashed line: the same, without
granularity. Dotted line: mixture of trions and electrons ($f
=0.3$) . Thin solid line: pure electron gas ($f = 0$). Distance
between impurity and carrier planes: $d = 4 \text{nm}$. Other
parameters as in Fig. \ref{fig_q_fraction}.
\label{fig_scattering_collage}}
\end{figure}

The scattering induced by charged impurities is the main
interaction process affecting the transport properties of charge
carriers at low temperatures. Within the Fermi's golden rule, the
wave-vector dependent lifetime of trions $\tau_{\text{tr}}(k)$ in
presence of the mixed gas of trions and electrons is given by the
expression
\begin{equation}
\frac{1}{\tau_{\text{tr}}(k)}  = \frac{N_\text{imp}}{A} \left (
\frac{2\pi M e^4}{\kappa^2 \hbar^3} \right ) \int_{0}^{2\pi}
d\theta \frac{\exp{(-2dq)}}{(q \epsilon_{\text{mix}}(q,0))^2}~,
\label{lifetime}
\end{equation}
with $q = 2k~|\sin{(\theta/2)}|$ and $N_{\text{imp}}/A$ is the
density of impurities per unit area. Notice that the electron
lifetime is $\tau_{\text{e}}(k) = (M/m_e) \tau_{\text{tr}}(k)$.
The velocity lifetime $\tau^{\text{v}}_{\text{tr}}(k)$ is given by
same expression in Eq. (\ref{lifetime}), but with an additional
factor $(1-\cos{\theta})$ within the integral. In Fig.
\ref{fig_scattering_collage}, we plot the scattering integral
$I_{\text{scatt}}(k) = \int_{0}^{2\pi} d\theta \exp{(-2dq)}/(q
\epsilon_{\text{mix}}(q,0))^2$ (in units of $a^{*2}$). In the case
of a pure trion gas ($f=1$, thick solid line), the $k$-dependence
of the scattering integral has a resonant structure. The peak is
due to the node of the granularity factor $|\mathcal{T}(q)|^2$
(see Fig. \ref{fig_granularity}). Indeed, around the nodal
wave-vector, the trion screening is dramatically quenched and the
scattering efficiency enhanced. This effect is absent if the
granularity of the trion is not taken into account (thick dashed
line) and in the case of a pure electron gas (thin solid line).
Note that in the case of a mixture of trions and electrons, the
contribution of trions is dominant, as shown by the dotted line,
corresponding to a mixture where the trion fraction is only $30
\%$. The results for the velocity lifetimes (not shown) are
qualitatively analogous.

Similarly to the case of electrons, a pure gas of trions has
longitudinal collective excitations (trionic plasmons), which are
given by the solution of the equation $\epsilon_{X^-}(q,\omega(q))
= 0$. In the long wave-length limit, the trionic plasmons behave
exactly as the electronic plasmons \cite{Ando_RMP}, once the
electron mass $m_e$ is replaced by the trion mass $M = 2m_e+m_h$.
For finite wave-vectors, as a result of the trion granularity, the
plasmon frequency is decreased, due to the quenching of the
screening efficiency. For a finite wave-vector $q$, when only the
trion ground state is involved, the behavior is that of an
electron gas, once the electron charge $e$ is replaced by the
effective charge $e'(q)= e |{\mathcal T(q)}| < e$, as it can be
deduced from Eqs. (\ref{epsilon_trion}) and (\ref{bubble_trion}).
For excitation energies of the order of the trion binding energy,
the excited states will give extra plasmon branches corresponding
to transition between internal states. For a mixed gas of trions
and electrons, the plasmon branches are the solutions of
$\epsilon_{mix}(q,\omega(q)) = \epsilon_{X^-}(q,\omega(q)) +
\epsilon_{e^-}(q,\omega(q)) -1 = 0$. Hence, a coupling is present
between the trionic and the electronic branches of plasmons. The
complex phenomenology of the collective excitations in presence of
a mixture of electrons and trions will be addressed in detail in a
future publication.

In conclusion, we have investigated the screening response of
trions, taking into account their composite nature. We have
obtained a RPA-dielectric function for a mixed gas of trions and
electrons. This interesting physical regime is experimentally
achievable by resonant optical pumping of trions in doped QWs. In
the static regime, we have shown the major impact of the trion
granularity in determining the dielectric response at finite
wave-vectors. We have calculated the scattering rates of electrons
and trions due to the interaction with the charged impurities. The
internal motion of trions is responsible for a quenching of the
screening response for wave-lengths comparable or smaller than the
trion effective radius. Moreover, the granularity produces
resonant features in the wave-vector dependence, due to the
compensation which can occur between the contribution of the two
electrons and the hole within the $X^-$. We hope that our study
will stimulate the research in the fundamental properties of
trions in the dense regime. We expect that the effect of trion
granularity in the screening response will have an impact on the
transport properties of these charged particles, whose
investigation is underway \cite{Sanvitto_2001,Pulizzi_2003}. In
particular, the impurity-induced localization and the onset of the
metal-insulator transition\cite{Eytan_1998} should be considerably
modified by the conversion of electrons into trions through
resonant optical pumping.

LPA-ENS (former LPMC-ENS) is "Unit\'{e} Mixte de Recherche
Associ\'{e} au CNRS (UMR 8551) et aux Universit\'{e}s Paris 6 et
7".

\end{document}